\documentclass[11pt]{article}
\usepackage{amssymb}

\newcommand{\beq}{\begin{equation}}
\newcommand{\eeq}{\end{equation}}
\newcommand{\be}{\begin{equation}}
\newcommand{\ee}{\end{equation}}

\def \e  {\mathop{\rm e}\nolimits}
\def \be  {\begin{equation}}
\def \ee  {\end{equation}}
\def \ba  {\begin{eqnarray}}
\def \ea  {\end{eqnarray}}

\def\be{\begin{equation}}
\def\ee{\end{equation}}
\def\bea{\begin{eqnarray}}
\def\eea{\end{eqnarray}}
\newcommand \ci [1] {\cite{#1}}
\newcommand \bi [1] {\bibitem{#1}}
\def \lab #1 {\label{#1}}
\newcommand\re[1]{(\ref{#1})}

\newcommand\lr[1]{{\left({#1}\right)}}

\def \Tr {\mbox{Tr\,}}

\newcommand \vev [1] {\langle{#1}\rangle}

\newcommand \ket [1] {|{#1}\rangle}

\def \CO {{\cal O}}
\font\cmss=cmss10 \font\cmsss=cmss10 at 7pt
\def\inbar{\,\vrule height1.5ex width.4pt depth0pt}
\def\IC{\relax\hbox{$\inbar\kern-.3em{\rm C}$}}
\def\IZ{\relax\ifmmode\mathchoice
{\hbox{\cmss Z\kern-.4em Z}}{\hbox{\cmss Z\kern-.4em Z}} {\lower.9pt\hbox{\cmsss
Z\kern-.4em Z}} {\lower1.2pt\hbox{\cmsss Z\kern-.4em Z}}\else{\cmss Z\kern-.4em
Z}\fi}
\def\IR{{\hbox{{\rm I}\kern-.2em\hbox{\rm R}}}}
\def\IP{{\hbox{{\rm I}\kern-.2em\hbox{\rm P}}}}
\newcommand{\as}{\ifmmode\alpha_{\rm s}\else{$\alpha_{\rm s}$}\fi}
\renewcommand{\Re}{\mathop{\rm Re}\nolimits}

\newcommand \Mybf[1] {\mbox{\boldmath$ {#1} $}}
\newcommand \mybf[1] {\mbox{\boldmath$ {\scriptstyle #1} $}}

\begin{document}

\setcounter{footnote}{0}

\setcounter{equation}{0}
\centerline{
\bf String/Gauge Correspondence; View from
the High Energy Side
}
\vspace{0.3cm}
\centerline{ A. Gorsky}

\begin{center}
{\em Institute of Theoretical and Experimental Physics, \\
B.Cheremushkinskaya 25, Moscow,  117259, Russia }

\end{center}
\begin{abstract}
We briefly review the recent progress concerning the application
of the hidden integrability to the derivation of the
stringy/brane picture for the high energy QCD.
\end{abstract}

\section{Introduction}
 The explicit realization of the generic string/gauge correspondence
program (see,
for instance, \cite{polyakov})
remains the challenging problem during the last decades. It escaped the
complete solution apart from the simplified two-dimensional
example \cite{gt}. Some time ago it was attacked in the SUGRA
approximation at the string side \cite{maldacena} where the
extended N=4 SUSY in gauge theory allowed to identify
the dual classical SUGRA solution. More recently there was some
progress beyond SUGRA approximation \cite{bmn} where it was
found that the string theory in the Penrose limit of $AdS_5 \times S^5$
background corresponds to some sector of the N=4 SYM theory.
Moreover, it was argued \cite{gkp} that the classical solution
in the $\sigma$ model corresponding to the rotating long string
amounts to the anomalous dimensions of the operators extended
along the light-cone.

In the recent paper \cite{gkk} which this talk is mainly
based on it was suggested that
a kind of stringy picture can be developed in another
regimes, namely in the Regge and light-cone kinematics
for a non supersymmetric gauge theory.
This hypothetical string theory in principle should
be responsible for the Veneziano type amplitudes
in QCD.
Since the theory
is not supersymmetric one can not expect any conformal
symmetry in the low energy sector. However, there is
a remnant of the conformal symmetry in the high energy sector.
Actually, it was exploited long time ago in calculations
of the anomalous dimensions in QCD \cite{mak}. The
symmetry which substitutes SO(2,4) conformal group
in d=4 is SL(2,C) in the Regge limit and collinear
conformal group SL(2,R) on the light-cone. According to
the string/gauge correspondence these groups of the
global symmetry should be related to the isometry
of the gravity background. Hence the natural  gravity
background to be expected should involve Euclidean $AdS_3$ factor for the Regge case
and $AdS_2$ on the light-cone. Note that our approach
differs from the one considered in
\cite{polchinski,janik2,giddings} to describe
high energy amplitudes in N=4 SYM theory.

It appears that the place where the stringy picture
can be captured is the RG evolution. It is the RG flow
where the relevant symmetry can be traced out. The
key property of RG flows to be used is the hidden
integrability which can be identified in both
regimes under consideration. To define the relevant
integrable system we have to extract the proper
degrees of freedom, set of integrals of motion
and the corresponding "times". The degrees of freedom
in the Regge case have been identified as the
coordinates of the reggeized gluons \cite{lipatov,fk}
while in the light-cone kinematics these are
fields involved in the bare Lagrangians like quarks
and gluons \cite{BDM,bm,b,bk}. The time $t_0$
in the integrable system has the meaning of some
evolution parameter; namely, $log s$ in Regge case and
$log \mu$ in the light-cone case. The evolution
equations can be presented in the Hamiltonian form
where the anomalous dimensions for the light-cone
operators  or intersepts
of the multireggeon states in the Regge case
play the role of the eigenvalues of integrable spin chain
Hamiltonians.

On the other hand, there is another class of (supersymmetric) Yang-Mills theories
in which integrability emerges. In the $\cal{N}$=2 SUSY YM theory which can be solved
in the low-energy limit exactly \cite{sw1} the relevant integrable system was
identified as the complexified periodic Toda chain. The low-energy effective
action and the BPS spectrum in the theory can be described in terms of the
classical Toda system \cite{gkmmm}. In the relevant case of
the superconformal $N_f=2N_c$ theory, the solution leads to the periodic XXX
chain in the magnetic field~\cite{ggm}. Because these magnets are classical, the
spins are not quantized and their values are determined by the matter masses. In
the special case, when the matter is massless, one arrives at the classical XXX
magnet of spin zero.

At the first glance there is nothing in common neither between high-energy
(Regge) limit of QCD and low-energy limit of SUSY YM, nor between the
corresponding integrable models. However it
appears that there is a deep relation between the Regge limit of QCD and some
particular limit of superconformal $\cal{N}$=2 SUSY YM at $N_f=2N_c$. This
relation is based on the fact that in the both cases we are dealing with the same
class of integrable models -- the $SL(2,\mathbb{C})$ homogenous spin zero
Heisenberg magnets. There is the key difference between two theories;
the integrable model in the SUSY YM is a classical Heisenberg magnet, whereas the
Regge limit of QCD is described by a quantum Heisenberg magnet. To bring two
pictures together, it is natural to develop quasiclassical description in QCD
case and quantize the classical picture in SUSY YM.

In \cite{gkk} we  developed a  brane picture for the Regge
limit of the multi-colour QCD. We  argue that the dynamics of the
multi-reggeon compound states in multi-colour QCD is described in this picture by
a single membrane wrapped around the spectral curve of the Heisenberg magnet. The
fact that we are dealing with the {\it quantum\/} magnet implies that the moduli
of the complex structure of the spectral curve can take only quantized values.
Let us note that in the
Regge limit there is a natural splitting of the four-momenta of reggeized gluons
into two transverse and two longitudinal components, which is unambiguously
determined by the kinematics of the scattering process. One can make a Fourier
transform with respect to the transverse momenta and define a mixed
representation: two-dimensional longitudinal momentum space and two-dimensional
impact parameter space for which it is natural to use complex coordinates $z$ and
$\bar z=z^*$. The Riemann surface which the brane is wrapped around lives in this
``mixed'' coordinate-momenta space. Different projections of the wrapped brane
represent hadrons and reggeons. The genus of this surface is fixed by the number
of reggeons. Let us note that multi-reggeon states appear from the summation of
only planar diagrams in QCD and from the point of view of topological expansion
they all correspond to the cylinder-like diagram in a colour space. It
contribution to the scattering amplitude is given by the sum of $N-$reggeon
exchanges in  the $t-$channel with $N=2,3...$. It is amusing that in result we
get a picture where we sum over Riemann surfaces of arbitrary genus $g=N-2$.

\section{Riemann surfaces and universality class of the Regge regime}
In the Regge limit the two particle scattering can be described
in terms of the reggeon degrees of freedom.
The calculation of the $N-$reggeon diagrams can be performed using the
Bartels--Kwiecinski--Praszalowicz approach \ci{Bar}. The diagrams
have a generalized ladder form and for fixed number of reggeons, $N$,
their contribution satisfies Bethe--Salpeter like equations for the scattering of
$N$ particles. The solutions to these equations define the colour-singlet
compound states $\ket{\Psi_N}$ built from $N$ reggeons. These states propagate in
the $t-$channel between two scattered onia states and give rise to the following
(Regge like) high energy asymptotics of the onium-onium cross-section
\be
\sigma_{\rm tot}(s) \sim \sum _{N=2\,,3,...} (\as N_c)^N \frac{x^{-{\as}N_c \varepsilon_N/\pi}}
{\sqrt{\as N_c\ln 1/x}}\,\beta_A^{(N)}(Q^2)\beta_B^{(N)}(Q^2)
\label{amp}
\ee
with the exponents $\varepsilon_N$ defined below. Here, each term in the sum is
associated with the contribution of the $N-$reggeon compound states. At $N=2$,
the corresponding state defines the BFKL pomeron \cite{BFKL} with
$\varepsilon_2=4\ln 2$.

The $N-$reggeon states are defined in QCD as solutions to the Schr\"odinger
equation
\be
{\cal H}_N \Psi(\vec z_1,\vec z_2,...,\vec z_N) = \frac{\as}{\pi}
N_c\varepsilon_N
\Psi(\vec z_1,\vec z_2,...,\vec z_N)
\label{Sch}
\ee
with the effective QCD Hamiltonian ${\cal H}_N$ acting on two-dimensional
transverse coordinates of reggeons, $\vec z_k$  ($k=1,...,N$) and their colour
$SU(N_c)$ charges
\be
{\cal H}_N=-\frac{\as}{2\pi} \sum_{1 \le i < j \le N} H_{ij} \ t_i^a t_j^a \,.
\lab{ham}
\ee
Here, the sum goes over all pairs of reggeons.

To get some insight into the properties of the $N-$reggeon states, it proves
convenient to interpret the Feynman diagrams as describing a
quantum-mechanical evolution of the system of $N$ particles in the $t-$channel
between two onia states $\ket{A}$ and $\ket{B}$
\be
\sigma_{\rm tot}(s)=\sum_{N\ge 2} (\as N_c)^N \vev{A|\e^{\ln (1/x) \cdot{\cal H}_N}|B}\,,
\label{time}
\ee\\[-2mm]
with the rapidity $\ln x=\ln Q^2/s$ serving as Euclidean evolution time. To find
the high energy asymptotics of \re{time}, one has to solve the Schr\"odinger
equation \re{Sch} for arbitrary number of reggeized gluons $N$, expand the
operator $(1/x)^{{\cal H}_N}$ over the complete set of the eigenstates of ${\cal
H}_N$ and, then, resum their contribution for arbitrary $N$.

 The pair-wise interaction between reggeons
is reduced in the multi-colour limit to a nearest-neighbor
interaction  and, in addition, the colour operator is replaced by a
$c-$valued factor $t^a_it^a_j \to -N_c/2$. The
two-particle reggeon Hamiltonian $H_{i,i+1}$ acts on the two-dimensional
(transverse) reggeon coordinates $\vec z=(x,y)$ and it coincides with the BFKL
kernel. It becomes convenient to introduce the complex valued
(anti)holomorphic coordinates, $z=x+iy$ and $\bar z=x-iy$, and parameterize the
position of the $k$th reggeon as $\vec z_k=(z_k,\bar z_k)$. One can rewrite the $N-$reggeon
Hamiltonian  as
\be
{\cal H}_N = \frac{\as N_c}{4\pi}\lr{H_N + \bar H_N} + \CO(N_c^{-2})\,.
\lab{RH}
\ee
Here the hamiltonians $H_N$ and $\bar H_N$ act on the (anti)holomorphic
coordinates and describe the nearest--neighbour interaction between $N$ reggeons
$$
H_N = \sum_{m=1}^N H(z_m,z_{m+1})\,,\qquad
\bar H_N = \sum_{m=1}^N H(\bar z_m,\bar z_{m+1})\,,
$$
with the periodic boundary conditions  $z_{k+1}=z_1$ and $\bar z_{k+1}=\bar z_1$.
The interaction Hamilonian between two reggeons with the coordinates $(z_1,\bar
z_1)$ and $(z_2,\bar z_2)$ in the impact parameter space, is given by the BFKL
kernel
\be
H(z_1,z_2)=-\psi(J_{12})-\psi(1-J_{12})+2\psi(1)\,,
\lab{psi}
\ee
where $\psi(x)=d\ln\Gamma(x)/{dx}$ and the operator $J_{12}$ is defined as a
solution to the equation
\be
J_{12} (J_{12}-1) = -(z_1-z_2)^2\partial_1\partial_2
\lab{J}
\ee
with $\partial_m=\partial/\partial z_m$. The expression for $H(\bar z_1,\bar
z_2)$ is obtained from \re{psi} by substituting $z_k\to\bar z_k$.

In the quasiclassical limit the solutions to the equations
of motion of the spin chains can be described in terms of the Riemann
surfaces \cite{kk} and we will argue that just these surfaces
fix the universality class of the Regge regime. Moreover it
appears that it falls into the same universality class
as superconformal N=2 SQCD at the strong coupling orbifold point.

Let us recall the main features of the low-energy effective action in the
$\cal{N}$=2 SUSY YM theories relevant for our purposes \cite{sw1,sw2}. The key
point is that the theory has the nontrivial vacuum manifold since the potential
involves the term
\beq
V(\phi)= \Tr [\phi,\phi^{+}]^2\,.
\eeq
Here $\phi$ is the complex scalar field which generically develops the vacuum
expectation value
\beq
\phi={\rm diag}(\phi_1,....,\phi_{N_c})\,,
\eeq
with $\Tr \phi=0$. The gauge invariant order parameters $u_k=\vev{0|\Tr\phi
^k|0}$ parameterize the Coulomb branch of the vacuum manifold. They define a
scale in the theory with respect to which one can discuss the issue of a low
energy effective action. This action takes into account one-loop perturbative
correction as well as the whole instanton series and it is governed by the
Riemann surface of the genus $N_c-1$ for $SU(N_c)$ gauge group. The same Riemann
surface appears as the spectral curve of a classical integrable many-body system
(see \cite{gm} for a review).
The integrable system provides the natural explanation
for the appearance of the meromorphic differential $\lambda_{\rm SW}$, which
turns out to be the action differential in the separated variables
 $\lambda_{_{\rm SW}}= p\,dx$. Let us emphasize that  for SUSY
YM case the $\it{classical}$ integrable system is relevant  and the meaning of
the quantum system and the corresponding spectrum for SUSY YM remains an open
question. It should involve the quantization of the vacuum moduli space  since
the hamiltonian in the dynamical system is nothing but $H = u_2 =\vev{\Tr\Phi
^2}$.
Simultaneously, this parameter serves as the coordinate on the moduli space
of the complex structures of the Riemann surfaces whiich means that
 the quantization of the
integrable system is related with the
quantization of the effective d=2 gravity.

Let us now consider a particular theory, namely the superconformal $\cal{N}$=2
SUSY YM with $N_f=2N_c$ massless fundamental hypermultiplets \cite{sw2,super}.
The corresponding integrable system is described by the spectral curve
$\Sigma_{N_c}$
\beq
 y^2=P_{N_c}^2(x) -4x^{2N_c}(1- \rho^2(\tau_{cl}))\,,
\label{SUSY-curve}
\eeq
where $\rho^2(\tau_{cl})$ is some function of a coupling constant of the theory
and the polynomial $P_{N_c}$ depends on the coordinates
on the moduli space $\vec u=(u_2,...,u_{N_c})$
\be
P_{N_c}(x)= \sum_{k=0}^{N_c} q_k(\vec u)\, x^{N_c-k} =2x^{N_c}+ q_2\, x^{N_c-2} +
... + q_{N_c}\,,
\ee
where $q_0=2$, $q_1=0$ and other $q_k$ are some known functions of $\vec u$.
Their explicit form is not important for our purposes. The Seiberg-Witten
meromorphic differential on the curve is given by
\be
\lambda_{_{\rm SW}}= p\,dx= \ln({\omega}/{x^{N_c}})\,dx\,,
\label{lambda-SW}
\ee
where $y=\omega-x^{2N_c}/\omega$.

{}From the point of view of integrable models the spectral curve
corresponds to a classical XXX Heisenberg spin chain of length $N_c$ with the
spin zero at all sites (due to $q_1=0$) and parameter $\rho$ related to the
external magnetic field, or equivalently, to the twisted boundary conditions
\cite{ggm}.

Comparing the spectral curve  for the superconformal
$\cal{N}$=2 SUSY YM with $N_f=2N_c$ with the spectral curve  for
$N-$reggeon compound states in multi-colour QCD one   observes that
they coincide if we make the following identification
\begin{itemize}
\item The number of the reggeons $N=N_c$;
\item The integrals of motion of multi-reggeon system are identified as the above
mentioned functions $q_k(\vec u)$ on the moduli space of the superconformal
theory;
\item The coupling constant of the gauge theory should be such that
$\rho(\tau_{cl})=0$.
\end{itemize}
Under these three conditions the both theories fall into the same universality
class.
The third condition implies that the so-called strong
coupling orbifold point on the moduli space corresponds to the Regge limit of
multi-colour QCD.
For instance, the strong coupling orbifold point $\rho(\tau_{\rm cl})=0$ describing the Odderon
state in QCD occurs at
\beq
\tau_{\rm cl}=\frac12 +\frac{i}{2\sqrt{3}}\,.
\label{tau-N=3}
\eeq

Finally we would like to note that there is the following intriguing fact. In the
case of $N_c=2$ which corresponds on the QCD side to the BFKL Pomeron state, the
effective coupling constant is given in the weak coupling regime by the
expression
\be
\tau_{\rm eff}= \tau_{\rm cl} + i\frac{4\ln 2}{\pi}
+ \sum_{k}c_k\cdot \e^{2i\pi k\tau_{\rm cl}}, ~~~
\ee
where the second term  is due to a finite one-loop
correction~\cite{khoze} and the rest is the sum of instanton contributions. It is
amusing that this one-loop correction to the coupling constant
\be
\frac{1}{g^2_{\rm eff}}=
\frac{1}{g^2_{\rm cl}}\left[1 + \frac{g^2_{\rm cl}}{4\pi^2} 4\ln 2\right] +...
\ee
 coincides with the expression for the
intercept of the BFKL Pomeron  after one identifies bare coupling
constant in the superconformal
theory with the t'Hooft coupling constant in QCD.

\section{Brane picture for the Regge limit}

Let us discuss now using the universality class
arguments above  a stringy/brane picture for the Regge limit in multi-colour
QCD.
To warm up we would like to recall the brane description of the low-energy
dynamics of the ${\cal N}=2$ SUSY YM. In the IIA framework the pure gauge theory
is defined on the worldvolume of $N_c$ D4 branes with the coordinates
$(x_0,x_1,x_2,x_3,x_6)$ stretched between two NS5 branes with the coordinates
$(x_0,x_1,x_2,x_3,x_4,x_5)$ and displaced along the coordinate $x_6$ by an amount
inversely proportional to the coupling constant, $\delta x_6=1/g^2$
\cite{witten}. The coordinates at which the D4 branes intersect with the
$(x_4,x_5)$ complex plane define the vacuum expectation value of the scalar
fields. This picture agrees perfectly with the RG behaviour of the coupling
constant and yields the correct beta function in the gauge theory. The Riemann
surface $\Sigma$ discussed above describes the vacuum state of the theory and the
spectrum of the stable BPS states. The lifting to the M theory picture leads to
emergence of a single M5 brane with the worldvolume $R^4\times\Sigma$ \cite{warner}.

In our case, we also need to incorporate into this picture branes corresponding
to the fundamental matter with $N_f=2N_c$. There are two ways to do this: either
using semi-infinite D4 branes lifted into M5 brane in the M theory, or using D6
branes which induce  the nontrivial KK monopole background for the M2
brane wrapped on  the Riemann surface \cite{witten}.
As was shown in \cite{ggm} in the latter case the resulting
brane picture remains consistent with the integrable spin chain dynamics and we
shall stick to this case.

The explicit metric of the KK background in the M theory involving
$(x_4,x_5,x_6,x_{10})$ coordinates has the multi-Taub-NUT form
\beq
ds^2= \frac{V}{4}d\vec{r}^2 + \frac{V^{-1}}{4}(d\tau +\vec{A}d\vec{r})^2
\eeq
where $\vec{r}=(x_4,x_5,x_6)$,~ $\tau =x_{10}$ and
$\vec{A}$ is the Dirac monopole potential. The magnetic charge
comes  from the nontrivial twisting of $S^1$ bundle over $R^3$.
The function $V$ behaves as
\beq
V= 1 + \sum_{i=1}^{i=N_f} \frac{1}{|\vec{r} - \vec {r_i}|}
\eeq
where $\vec r_i = (x^i_4,x^i_5,x^i_6)$ are the positions of six-branes. For a
superconformal case one must have  $x^i_4 = x^i_5 =0$ and positions of sixbranes
in $x_6$ direction are irrelevant.


Let us turn now to our proposal for
the brane realization of the Regge limit. We
shall explore the brane representations for the $N_f=2N_c$ theory known in the
IIA/M theory \cite{witten}. However unlike the SUSY
case where the spectral curve is
embedded in the internal ``momentum'' space the
spectral curve of the noncompact spin chain is placed in the phase space
involving both impact parameter plane as well as momenta.
Consider the IIA/M type picture which is reminiscent to the
realization of SYM theory via two NS5 and $N_c$ D4 branes. We suggest that the
coordinates involved in ``IIA'' picture are the transverse impact parameter
coordinates $x_1,x_2$ and rapidity $\lambda=\ln k_{+}/k_{-}$. Transverse
coordinates are analogue of $(x_4,x_5)$  coordinates in SUSY case while rapidity
substitutes the $x_6$ coordinate.
Now let us make the next step and suggest that similar to SUSY case the single
brane is wrapped around the spectral curve of XXX magnet and two ``hadronic
planes" together with N ``Reggeonic strings'' are just different projections of
the single membrane with worldvolume $R\times \Sigma$. The coordinates involved
into configurations are $x_1+i x_2$ and $y=e^{-(\lambda +ix_{10})}$ where
$x_{10}$ is the ``M-theory'' coordinate.

Let us emphasize once again that
the brane configuration for Regge limit contrary
to SYM case partially involves
the coordinate space. More precisely the geometry
of $N_f=2N_c$ theory is determined by the following parameter \cite{witten}
\be
\xi = - \frac{4\lambda_{+}\lambda_{-}}{(\lambda_{+}-\lambda_{-})^2}\,.
\ee
Here $\lambda_{+}$ and $\lambda_{-}$ are asymptotic positions of five-branes
defined by the large $x$ behaviour of the curve
\beq
\omega \propto \lambda_{\pm} x^{N_c}\,.
\eeq
$\lambda_\pm$ can be found as roots of the equation
\beq
\lambda_\pm^2 +\lambda_\pm +\frac{1}{4}(1-\rho^2)=0
\eeq
Since Regge limit corresponds to the strong coupling orbifold point, $\rho=0$,
the value of $\xi$ is fixed as $\xi=\infty$. This corresponds to the coinciding
branes at infinity.

Finally, the M theory brane picture for the Regge limit involves M5 brane
corresponding to the vacuum state of the QCD. We can not say how it is placed
precisely as the minimal surface in the internal seven dimensional space since
the corresponding geometry is unknown yet. The new ingredient - membrane share
the time direction with the M5 brane and wrapping around the Riemann surface
which is embedded into two-dimensional complex ``phase space'' with the
multi-Taub-NUT metric determined by KK monopoles with the magnetic charge $2N$,
which is a double number of reggeized gluons participating in the scattering
process. The possible identification of the membrane above with the
M2 brane deserves further investigation.


\section{Quantum spectrum and S-duality}

The $S-$duality is a powerful symmetry in the SUSY YM theory which allows us to
connect the weak and strong coupling regimes. The effective coupling in this
theory coincides with the modular parameter of the spectral curve of the
underlying classical integrable model. As a consequence, the $S-$duality
transformations in the gauge theory are translated into the modular
transformations of the spectral curve describing complexified integrable system.
A formulation of the $S-$duality in the latter system naturally leads to an
introduction of the notion of the dual action~\cite{fgnr}.

So far the $S-$duality was well understood only for classical integrable models.
In the case of multi-colour QCD in the Regge limit the situation is more
complicated since the duality has to be formulated for a quantum integrable
model. The integrals of motion take quantized set of values and the coordinates
on the moduli space are not continuous anymore. Therefore the question to be
answered is whether it is possible to formulate some duality transformations at
the quantum level.

To study this question let us propose the WKB quantization conditions which are
consistent with the duality properties of the complexified dynamical system whose
solution to the classical equations of motion are described by the Riemann
surface. We recall that the standard WKB quantization conditions involve the real
slices of the spectral curve
\beq
\oint_{A_i}p\,dx = 2\pi \hbar (n_i+1/2)
\eeq
where $n_i$ are integers and the cycles $A_i$ correspond to classically allowed
trajectories on the phase space of the system. In our case the coordinate $x$ is
complex and arbitrary point on the Riemann surface is classically allowed. As a
result the general classical motion involves both $A-$ and $B-$cycles on the
Riemann surface. This leads to the following generalized WKB quantization
conditions for actions and dual actions
\beq
\Re\oint_{A_i}p\,dx = \pi \hbar n_i\,, \qquad \Re\oint_{B_i}p\,dx = \pi\hbar
m_i\,,
\label{WKB-new}
\eeq
where the ``action'' differential was defined in \re{lambda-SW}. Note that in the
context of the SUSY YM this condition would correspond to the nontrivial
constraints on the periods  and on the mass spectrum of the BPS
particles . It is clear that the WKB conditions \re{WKB-new} imply
the duality $A_i\leftrightarrow B_i$ and $n_i\leftrightarrow m_i$.

Let us consider the quantization conditions \re{WKB-new} in the simple case of
the Odderon $N=3$ system. The spectral curve  is a torus
\beq
y^2=(2x^3+q_2x +q_3)^2 - 4x^6
\eeq
where $q_2$ is given by conformal spin while $q_3$ is the complex integral of
motion to be quantized. The quantization conditions \re{WKB-new} read (for $\hbar
=1$)
\beq
\Re a(q_3) = \pi n, \qquad \Re a_D(q_3) = \pi m
\label{a-N=3}
\eeq
where $n$ and $m$ are integer. These equations can be solved for large values of
$q_3^2/q_2^3 \gg 1$, for which the expressions for the periods $a(q_3)$ and
$a_{D}(q_3)$ are simplified considerably. The explicit evaluation of the
integrals  in this limit  yields
\be
a(q_3) = \frac{(2\pi)^2 q_3^{1/3}}{\Gamma^3(2/3)}\,,\qquad a_D(q_3)
=a(q_3)\lr{\frac12+\frac{i}{2\sqrt{3}}}
\ee
Substituting these expressions into \re{a-N=3} one finds
\beq
q_3^{1/3}=\frac{\Gamma^3(2/3)}{2\pi}\left(\frac{\ell_1}2+i\frac{\sqrt{3}}2
\ell_2\right)\,.
\label{q3-quan}
\eeq
where $\ell_1=n$ and $\ell_2=n-2m$.
The WKB expressions \re{q3-quan} are in a good agreement with the exact
expressions for quantized $q_3$ obtained from the numerical solutions of the Baxter
equations in \cite{kornew}. Note that WKB formulae can not be naively
applied to the ground states discussed in \cite{kornew,janik1,devega}.

Let us emphasize that the point at the moduli space corresponding to the
degeneration of the torus for the Odderon case does not appear in the quantum
spectrum.  From the point of view of the gauge theory this means that the
appearance of the massless states is forbidden.

In the general multi-reggeon case we have to consider the quantization conditions
\re{WKB-new} on the Riemann surface  of the genus $(N-2)$ which has
the same number of the $A-$ and $B-$cycles. In result the spectrum of the
integrals of motion $q_3$, $...$, $q_N$ is parametrized by two $(N-2)-$component
vectors $\vec{n}$ and $\vec{m}$. In the SUSY YM case these vectors define the
electric and magnetic charges of the BPS states. In the Regge case the physical
interpretation of $\vec{n}$ and $\vec{m}$ is much less evident. Let us first
compare the electric quantum numbers in the two cases. In Regge case it
corresponds to rotation in the coordinate space around the ends of the Reggeons.
This picture fits perfectly with the interpretation of the electric charge in
SUSY YM case. Indeed  VEVs
 of the complex scalar take values on  the  complex  plane which is the
counterpart of the impact parameter plane and
the rotation of the phase of the complex scalar is indeed the ``electric
rotation''.

To get some guess concerning the ``magnetic'' quantum numbers it is
instructive to check the geometrical picture behind them in the
simplified ``IIA'' picture.
 All  states corresponding to the ``electric'' degrees of freedom
are related to  fundamental strings  encircling  ``reggeoinic'' tubes
and  don't feel the
hadronic planes. However the ``magnetic'' states as is well known from SUSY YM
case are represented by the membrane stretched between two strings and two
hadronic planes.
Therefore  these states are sensitive to  hadronic  quantum numbers.
More detailed  interpretation of ``magnetic'' degrees of freedom in
the Regge regime  has to be recovered.

\section{Stringy/brane picture and the calculation
of the anomalous dimensions}

We have demonstrated that integrability properties of
the Schr\"odinger equation for the compound state of Reggeized gluons give rise
to the stringy/brane picture for the Regge limit in multi-colour QCD. There is
another limit in which QCD exhibit remarkable properties of integrability. It has
to do with the scaling dependence of the structure functions of deep inelastic
scattering and hadronic light-cone wave functions in QCD. In the both cases, the
problem can be studied using the Operator Product Expansion and it can be
reformulated as a problem of calculating the anomalous dimensions of the
composite operators of a definite twist. The operators of the lowest twist have
the following general form
\ba
&&{\cal O}^{(2)}_{N,k} (0) = (y D)^k \Phi_1(0) (y D)^{N-k}
\Phi_2 (0),
\nonumber
\\[2mm]
&&{\cal O}^{(3)}_{N,\mybf{k}}(0) =(y D)^{k_1} \Phi_1(0) (y D)^{k_2}
\Phi_2(0)(y D)^{N-k_1-k_2}\Phi_3(0),
\ea
where $\Mybf{k}\equiv (k_1,k_2)$ denotes the set of integer indices $k_i$,
$y_\mu$ is a light-cone vector such that $y_\mu^2=0$. $\Phi_k$ denotes elementary
fields in the underlying gauge theory and $D_\mu =
\partial_\mu -i A_\mu$ is a covariant derivative. The operators of a definite
twist mix under renormalization with each other. In order to find their scaling
dependence one has to diagonalize the corresponding matrix of the anomalous
dimension and construct linear combination of such operators, the so-called
conformal operators
\be
{\cal O}^{\rm conf}_{N,q}(0) = \sum_{\mybf{k}} C_{\mybf{k,q}}\cdot {\cal
O}_{N,\mybf{k}}(0)\,.
\label{conf-op}
\ee
A unique feature of these operators is that they have an autonomous RG evolution
\be
\Lambda^2\frac{d}{d \Lambda^2}\,{\cal O}^{\rm conf}_{N,q}(0)
=-\gamma_{N,q} \cdot{\cal O}^{\rm conf}_{N,q}(0)\,.
\ee
Here $\Lambda^2$ is a UV cut-off and $\gamma_{N,q}$ is the corresponding
anomalous dimension depending on some set of quantum numbers $\Mybf{q}$ to be
specified below. It turns out that the problem of calculating the spectrum of the
anomalous dimensions $\gamma_{N,q}$ to one-loop accuracy becomes equivalent to
solving the Schr\"odinger equation for the $SL(2,\mathbb{R})$ Heisenberg spin
magnet. The number of sites in the magnet is equal to the number of fields
entering into the operators under consideration.

To explain this correspondence it becomes convenient to introduce nonlocal
light-cone operators
\be
F(z_1,z_2)= \Phi_1(z_1y) \Phi_2(z_2y)\,,\qquad F(z_1,z_2,z_3)=
\Phi_1(z_1y) \Phi_2(z_2y) \Phi_3(z_3y)\,.
\label{F}
\ee
Here $y_\mu$ is a light-like vector $(y_\mu^2=0)$ defining certain direction on
the light-cone and the scalar variables $z_i$ serve as a coordinates of the
fields along this direction. The fields $\Phi_i(z_i y)$ are transformed under the
gauge transformations and it is tacitly assumed that the gauge invariance of the
nonlocal operators $F(z_i)$ is restored by including the Wilson lines between the
fields in the appropriate (fundamental or adjoint) representation. The conformal
operators appear in the OPE expansion of the nonlocal operators \re{F} for small
$z_1-z_3$ and $z_2-z_3$.

The field operators entering the definition of $F(z_i)$ are located on the
light-cone. This leads to the appearance of the additional light-cone
singularities. They modify the renormalization properties of the nonlocal
light-cone operators \re{F} and lead to nontrivial evolution equations which as
we will show below become related to integrable chain models. We notice that
there exists the following relation between the conformal three-particle
operators \re{conf-op} and the nonlocal operators \re{F}
\be
{\cal O}^{\rm conf}_{N,q}(0)=
\Psi_{N,q}(\partial_{z_1},\partial_{z_2},\partial_{z_3})
F(z_1,z_2,z_3)\bigg|_{z_i=0}\,,
\ee
where $\Psi_{N,q}(x_1,x_2,x_3)$ is a homogenous polynomial in $x_i$ of degree $N$
\be
\Psi_{N,q}(x_1,x_2,x_3)=\sum_{\mybf{k}} C_{\mybf{k,q}} \cdot x_1^{k_1} x_2^{k_2}
x_3^{N-k_1-k_2}
\ee
with the expansion coefficients $C_{\mybf{k,q}}$ defined in \re{conf-op}.
The problem of defining the conformal operators is reduced to finding the
polynomial coefficient functions $\Psi_{N,q}(x_i)$ and the corresponding
anomalous dimensions $\gamma_{N,q}$. Using the renormalization properties
 of the
nonlocal light-cone operators \re{F} one can show~\cite{bm}, that to
the one-loop accuracy the QCD evolution equation for the conformal
operators \re{F} can be rewritten in the form of a Schr\"odinger equation
\be
{\cal H}\cdot \Psi_{N,q}(x_i)
    =  \gamma_{N,q} \Psi_{N,q}(x_i)\,,
\label{gamma}
\ee
where the Hamiltonian ${\cal H}$ acts on the
$x_i-$variables which are conjugated
to the derivatives $\partial_{z_i}$ and, therefore, have the meaning of
light-cone projection $(y\cdot p_i)$ of the momenta $p_i$ carried by particles
described by fields $\Phi(z_i y)$.

For example  when $\Phi_1$ and $\Phi_2$ are quark fields of the same chirality
\be
F_{\alpha\beta}(z_1,z_2)=\sum_{i=1}^{N_c} (\bar
q_i^\uparrow{\not\hspace*{-0.59mm}y} )_\alpha(z_1y) ({\not\hspace*{-0.55mm} y}
q_i^\uparrow)_\beta(z_2y)
\ee
with $q_i^\uparrow=(1+\gamma_5)q_i/2$, the two-particle Hamiltonian is given
by
\begin{eqnarray}
  H_{12} = \frac{\as}{\pi}C_F \left[H_{qq}(J_{12})+1/4\right]\,,\qquad
  H_{qq}(J_{12})= \psi(J_{12})-\psi(2).
\label{H-qq}
\end{eqnarray}
where $C_F=(N_c^2-1)/(2N_c)$. The eigenfunctions for this Hamiltonian are the
highest weights of the discrete series representation of the $SL(2,R)$ group
\be
\Psi_{N}^{(2)}(x_1,x_2) = (x_1+x_2)^N C_N^{3/2} \left(\frac{x_1-x_2}{x_1+x_2}\right)
\ee
where $C_N^{3/2}$ are Gegenbauer polynomials. The corresponding eigenvalues
define the anomalous dimensions of the twist-2 mesonic operators built from two
quarks with the same helicity
\begin{eqnarray}
  \gamma_N^{(2)} = \frac{\as}{\pi}C_F[\psi(N+2)-\psi(2)+1/4]
  = \frac{\as}{\pi}C_F\left[\sum_{k=1}^N\frac 1{k+1}+\frac14\right]\, .
\end{eqnarray}
At large $N$ this expression has well-known asymptotic behaviour
$\gamma_N^{(2)}\sim \as C_F/\pi\ln N$.

It is  conformal symmetry which dictates that the two-particle Hamiltonian is a
function of  the Casimir operator of the $SL(2,\mathbb{R})$ group, but it does
not fix this function. The fact that this function turns out to be the Euler
$\psi$-function  leads to a hidden integrability of the evolution equations for
anomalous dimensions of baryonic operators . Namely, for baryonic
operator built from three quark fields of the same chirality
\be
F_{\alpha\beta\gamma}(z_1,z_2,z_3)=\sum_{i,j,k=1}^{N_c}\epsilon_{ijk}({\not\hspace*{-0.55mm}
y} q_i^\uparrow)_\alpha(z_1y) ({\not\hspace*{-0.55mm} y}
q_j^\uparrow)_\beta(z_2y) ({\not\hspace*{-0.55mm} y} q_k^\uparrow)_\gamma(z_3y)
\ee
the evolution kernel is given by~\cite{BDM,bm}
\be
{\cal H}^{(3)}= \frac{\as}{2\pi}\left\{(1+1/N_c)\left[H_{qq}(J_{12})+
H_{qq}(J_{23})+ H_{qq}(J_{31})\right]+3C_F/2\right\}
\label{H-qqq}
\ee
with $H_{qq}$ given by \re{H-qq}.
The Schr\"odinger equation \re{gamma} with the
Hamiltonian defined in this way has a hidden integral of motion
\be
q=i\left(\partial_{x_1}-\partial_{x_2}\right)
\left(\partial_{x_2}-\partial_{x_3}\right)
\left(\partial_{x_3}-\partial_{x_1}\right) x_1 x_2 x_3
\label{Q}
\ee
and, therefore, it is completely integrable. Similar to the Regge case,
one can identify \re{H-qqq} as the Hamiltonian of a quantum XXX Heisenberg magnet
of $SL(2,\mathbb{R})$ spin $j_q=1$.
The number of sites is equal to the number of quarks.

Based on this identification we shall argue now that the calculation of the
anomalous dimensions can be formulated entirely in terms of Riemann surfaces
which in turn leads to a stringy/brane picture. It is important to stress here
the key difference between Regge and light-cone limits of QCD.
In the first case the impact parameter space provides the complex plane for
the Reggeon coordinates and we are dealing with a $(2+1)-$dimensional
dynamical system. In the second case the QCD evolution occurs along
the light-cone direction and is described by
a $(1+1)-$dimensional dynamical system. As a consequence, in these two cases we
have two different integrable magnets: the $SL(2,\mathbb{C})$ magnet for the
Regge limit and the $SL(2,\mathbb{R})$ magnet for the light-cone limit. The
evolution parameters (``time'' in the dynamical models) are also different: the
rapidity $\ln s$ for the Regge case and the RG scale $\ln\mu$ for the anomalous
dimensions of the conformal operators.

Our approach to calculation of the anomalous dimensions via Riemann surfaces
looks as follows. For concreteness, we shall concentrate on the evolution kernel
\re{H-qqq}. Similarly to the Regge case, one starts with the finite-gap solution
to the classical equation of motion of the underlying $SL(2,\mathbb{R})$ spin
chain and identifies the corresponding Riemann surface
\be
\omega - \frac{x^6}{\omega} = 2 x^3 -(N+2)(N+3) x + q\,,\qquad
\omega=x^3 \e^p
\ee
where $q$ is the above mentioned integral of motion \re{Q} and $N$ is the total
$SL(2,\mathbb{R})$ spin of the magnet,
or equivalently the number of derivatives
entering the definition of the conformal operator \re{conf-op}.
 Note that the  Riemann surface corresponding
 to  the three-quark operator  has genus $g=1$, while $g=0$
 for the twist 2 operators.

At the next step we quantize the Riemann surface in Sklyanin's approach \cite{sklyanin}. We replace
$p=i\partial/\partial x$ and impose the equation of the complex curve as the
operator annihilating the Baxter function
\be
\lr{\e^{i\partial/\partial x}+\e^{-i\partial/\partial x}}x^3 Q(x)=
\left[2 x^3 -(N+2)(N+3) x + q\right]Q(x)\,.
\label{curve-R}
\ee
This leads to the Baxter equation for the
Heisenberg $SL(2,\mathbb{R})$ magnet of spin $j=1$
\beq
(x+i)^3 Q(x+i)+(x-i)^3Q(x-i)=
\left[2 x^3 -(N+2)(N+3) x + q\right]Q(x)\,.
\label{bax}
\eeq
Similar to the Baxter equation in the Regge case, this equation does not have a
unique solution. To avoid this ambiguity one has to impose the additional
conditions that $Q(x)$ should be polynomial in $x$. This requirement leads to the
quantization of the integral of motion. The resulting polynomial solution
$Q=Q_q(x)$ has the meaning of the one-particle wave function in the separated
variables $x$ which in the case of the $SL(2,\mathbb{R})$ magnet take arbitrary
real values.

Given the polynomial solution to the Baxter equation \re{bax}, one can determine
the eigenspectrum of the Hamiltonian \re{H-qqq} and in result the anomalous
dimensions of the corresponding baryon operators
\beq
\gamma^{(3)}_{N,q} =\frac{\as}{2\pi}\left[(1+1/N_c){\cal E}_{N,q}+3C_F/2\right]\,,\qquad
{\cal E}_{N,q}=
i\frac{Q_q'(i)}{Q_q(i)} - i\frac{Q_q'(-i)}{Q_q(-i)}
\label{energy}
\eeq
parameterized by the eigenvalues of the integral of motion \re{Q} given by
\be
q = 
-i\frac{{Q_q}(i)-{Q_q}(-i)}{Q_q(0)}
\ee

The solution to the Baxter equation simplifies greatly in the quasiclassical
approximation which is controlled by the total $SL(2,\mathbb{R})$ spin of the
system $N$. For $N\gg 1$ the spectrum of the integral of motion $q$ is determined
by the WKB quantization condition~\cite{kk}
\be
\oint_A p\,dx =2\pi (n+1/2)+{\cal O}(1/N)
\label{WKB-R}
\ee
where $p$ was introduced in \re{curve-R}. Here integration goes over the
$A$-cycle on the Riemann surface defined by the spectral curve \re{curve-R}. This
cycle encircles the interval on the real $x-$axis on which $|\e^p|>1$. Solving
\re{WKB-R} one gets
\be
q=\pm \frac{N^3}{\sqrt{27}}
\left[1-3\left(n+\frac12\right)N^{-1} +\CO(N^{-2})\right]\,.
\label{sol-R}
\ee
Taking into account that in this
case $q_2=-(N+3)(N+2)$ and $q_3=q$ we conclude that for $N\to\infty$ the system
is approaching the Argyres-Douglas point. Note also that the WKB quantization
conditions \re{WKB-R} involve only the $A-$cycle on the Riemann surface and
unlike the Regge case there is no $S-$duality in the quantum spectrum in
the light-cone case.

Finally, the spectrum of the anomalous dimensions in the WKB approximation  is
given by~\cite{kk,bm}
\be
{\cal E}_{N,q} = 2\ln 2 - 6 + 6\gamma_E + 2 {\rm Re} \sum_{k=1}^3
\psi(1+i\delta_k) + \CO(N^{-6}),
\label{Energy}
\ee
where $\delta_k$ are defined as roots of the following cubic equation:
\be
2\delta_k^3 - (N+2)(N+3)\delta_k + q=0
\ee
and $q$ satisfies \re{sol-R}.

What can we learn about  stringy picture from this information about anomalous
dimensions? Let us remind that in the spirit  of  string/gauge fields
correspondence  the anomalous dimensions of gauge field theory
operators coincide with excitation energies of a string in some particular
background. An important lesson that we have learned from the analysis of
the two- and three-quark operators is that in the first case the
anomalous dimensions are uniquely specified by a {\it single\/} parameter
$N$ which define the total $SL(2,\mathbb{R})$ spin.
In the second case, a new quantum number emerges due to
the fact that the corresponding dynamical system is completely integrable. The
additional symmetry can be attributed to the operator $q$ defined in \re{Q}.
{}From the point of view of a classical dynamics this operator generates the
winding of a particle around the $A-$cycles on the spectral curve. Within the
string/gauge fields correspondence one expects to reproduce these properties
using a description in terms of the same string propagating in different
backgrounds. One is tempting to suggest that different properties of the
anomalous dimensions of the two- and three-particle operators
 should be attributed to different properties of the background.
In the case of the twist-2 the anomalous dimensions
depend on integer $N$ which in the classical system has an interpretation
of the total $SL(2,\mathbb{R})$ angular momentum of the system.
On the stringy side the same parameter has a natural interpretation as a
string angular momentum.

In our approach we have the following correspondence
\ba
\mbox{operator} &\Longleftrightarrow& \mbox{Riemann surface}
\nonumber
\\
\mbox{twist of the operator} &\Longleftrightarrow& \mbox{genus of
the Riemann surface}
\nonumber
\\
\mbox{calculation of the anomalous dimension}
&\Longleftrightarrow& \mbox{quantization of the Riemann surface}
\nonumber
\ea
It seems that the Riemann surfaces whose moduli (the integrals of motion of the
spin chain) define the anomalous dimensions of the corresponding operators
describe the $\sigma-$model solutions found in \cite{gkp}. The precise relation
between two approaches need to be clarified further.

As we have mentioned the quantization of the Riemann surface can be performed most
effectively in terms of the Baxter equation. It is worth noting that the solution
to the Baxter equation can be identified as a wave function of D0 brane
\cite{gnr}. Quantization conditions arise from the requirement for the wave
function the D0 brane probe in the background of the Riemann surface to be well
defined.

In the case of the light-cone composite operators we have to incorporate into the
stringy picture a new quantum number which is parameterized by an integer $n$,
Eq.~\re{sol-R}, i.e. string excitation spectrum has now different sectors
parameterized by this integer. The natural way to interpret these sectors is to
identify $n$ with the winding number of a closed string. The corresponding
background for such scenario is offered by the Riemann surface itself with the
string wrapped around the $A-$cycle. It is an interesting open question if one
can interpret a momentum in WKB quantization condition \re{WKB-R} as a momentum
of a string  $T$-dual to the string with the winding number $n$.

Since the spectrum of the anomalous dimensions in QCD coincides with the spectrum
of the $SL(2,\mathbb{R})$ spin chain Hamiltonian it would be interesting to
explore further the symbolic relation
\beq
H_{\rm string}\propto H_{\rm spin}
\eeq
where string propagates in the background determined by the Riemann surface of
the spin chain.
The possible link to the explanation of
this relation looks as follows.It is known that hamiltonian formulation of the spin chains is
closely related  to the Chern-Simons (CS) theory with the inserted Wilson lines.
The number of the sites in the spin chain $N$ corresponds to the number of the
Wilson lines. The gauge group in the CS theory is the symmetry group of the
magnet. To get CS action from the natural AdS geometry let us remind that
$AdS_3$ can be reformulated in terms of  $SL(2,C)$ CS indeed \cite{witten2}.
Hence it is natural
to assume that the spin chain corresponding to Regge limit describes the motion of $N$
degrees of freedom in $AdS_3$ space.
This issue will be discussed in more details elsewhere.

I am grateful to I. Kogan and G. Korchemsky for the collaboration. The work
was supported in part by grants  CRDF-RP2-2247, INTAS-00-334 and RFBR 01-01-00549.

,

\begin{thebibliography}{99}
\bibitem{polyakov}
A.Polyakov, hep-th/9711002, hep-th/9809057, hep-th/0110196
\bibitem{gt}
D.~J.~Gross and W.~I.~Taylor,
Nucl.\ Phys.\ B {\bf 400}, 181 (1993)
[arXiv:hep-th/9301068].
\bibitem{maldacena}
J.~Maldacena,
Adv.\ Theor.\ Math.\ Phys.\  {\bf 2} (1998) 231 [Int.\ J.\ Theor.\ Phys.\  {\bf
38} (1998) 1113]
[arXiv:hep-th/9711200]. \\
S.~S.~Gubser, I.~R.~Klebanov and A.~M.~Polyakov,
Phys.\ Lett.\ B {\bf 428} (1998) 105
[arXiv:hep-th/9802109].\\
E.~Witten,
Adv.\ Theor.\ Math.\ Phys.\  {\bf 2} (1998) 253 [arXiv:hep-th/9802150].
\bibitem{bmn}
D.~Berenstein, J.~M.~Maldacena and H.~Nastase,
JHEP {\bf 0204}, 013 (2002)
[arXiv:hep-th/0202021].
\bibitem{gkp}
S.~S.~Gubser, I.~R.~Klebanov and A.~M.~Polyakov,
Nucl.\ Phys.\ B {\bf 636} (2002) 99
[arXiv:hep-th/0204051]. \\
S.~Frolov and A.~A.~Tseytlin,
JHEP {\bf 0206} (2002) 007
[arXiv:hep-th/0204226].
\bibitem{gkk}
A.~Gorsky, I.~I.~Kogan and G.~Korchemsky,
JHEP {\bf 0205}, 053 (2002)
[arXiv:hep-th/0204183].
\bibitem{mak}
S.J.\ Brodsky et al., Phys.\ Lett.\  {\bf B91} (1980) 239;
                      Phys.\ Rev.\  {\bf D33} (1986) 1881;
\\
Yu.M.\ Makeenko, Sov.\ J.\ Nucl.\ Phys. {\bf 33} (1981) 440;
\\
Th.\ Ohrndorf, Nucl.\ Phys.\ {\bf {B198}} (1982) 26;

\bibitem{polchinski}
J.~Polchinski and M.~J.~Strassler,
Phys.\ Rev.\ Lett.\  {\bf 88} (2002) 031601
[arXiv:hep-th/0109174].\\
J.~Polchinski and L.~Susskind,
arXiv:hep-th/0112204.\\
R.~C.~Brower and C.~I.~Tan,
arXiv:hep-th/0207144.

\bibitem{janik2}
R.~A.~Janik and R.~Peschanski,
Nucl.\ Phys.\ B {\bf 625}, 279 (2002) [arXiv:hep-th/0110024].
\bibitem{giddings}
S.~B.~Giddings,
arXiv:hep-th/0203004.
\bibitem{lipatov}
L.~N.~Lipatov,
JETP Lett.\  {\bf 59} (1994) 596 [Pisma Zh.\ Eksp.\ Teor.\ Fiz.\  {\bf 59} (1994)
571] [arXiv:hep-th/9311037].
\bibitem{fk}
L.~D.~Faddeev and G.~P.~Korchemsky,
Phys.\ Lett.\ B {\bf 342}, 311 (1995) [arXiv:hep-th/9404173];
\\ G.~P.~Korchemsky,
Nucl.\ Phys.\ B {\bf 443} (1995) 255 [arXiv:hep-ph/9501232].
\bi{BDM}  V.M.~Braun, S.E.~Derkachov, A.N.~Manashov, Phys.\ Rev.\ Lett.\ \textbf{81} (1998) 2020
[arXiv:hep-ph/9805225].
\bibitem{bm}
V.~M.~Braun, S.~E.~Derkachov, G.~P.~Korchemsky and A.~N.~Manashov,
Nucl.\ Phys.\ B {\bf 553}, 355 (1999) [arXiv:hep-ph/9902375].
\bi{b}   A.V.~Belitsky, Phys.\ Lett.\ B \textbf{453} (1999) 59 [arXiv:hep-ph/9902361];
          Nucl.\ Phys.\ B \textbf{558} (1999) 259 [arXiv:hep-ph/9903512];
          Nucl.\ Phys.\ B \textbf{574} (2000) 407 [arXiv:hep-ph/9907420].
\bibitem{bk}
S.~E.~Derkachov, G.~P.~Korchemsky and A.~N.~Manashov,
Nucl.\ Phys.\ B {\bf 566} (2000) 203 [arXiv:hep-ph/9909539].
\bibitem{sw1}
N.~Seiberg and E.~Witten,
Nucl.\ Phys.\ B {\bf 426}, 19 (1994)
\bibitem{sw2}
N.~Seiberg and E.~Witten,
Nucl.\ Phys.\ B {\bf 431} (1994) 484 [arXiv:hep-th/9408099].
\bibitem{gkmmm}
A.~Gorsky, I.~Krichever, A.~Marshakov, A.~Mironov and A.~Morozov,
Phys.\ Lett.\ B {\bf 355} (1995) 466 [arXiv:hep-th/9505035].
\bibitem{ggm}
A.~Gorsky, S.~Gukov and A.~Mironov,
Nucl.\ Phys.\ B {\bf 517} (1998) 409 [arXiv:hep-th/9707120].
\bibitem{Bar}
J. Bartels, Nucl. Phys. B175 (1980) 365.\\
J. Kwiecinski and M. Praszalowicz, Phys. Lett. B94 (1980) 413.
\bibitem{BFKL}
E.A. Kuraev,  L.N. Lipatov and V.S. Fadin,
          Phys. Lett. B60 (1975) 50;
          Sov. Phys. JETP 44 (1976) 443; 45 (1977) 199;
\\        Ya.Ya. Balitsky and L.N. Lipatov, Sov. J. Nucl. Phys. 28 (1978) 822.
\bibitem{kk}
G.~P.~Korchemsky,
Nucl.\ Phys.\ B {\bf 462}, 333 (1996) [arXiv:hep-th/9508025]. \\
G.~P.~Korchemsky and I.~M.~Krichever,
Nucl.\ Phys.\ B {\bf 505}, 387 (1997) [arXiv:hep-th/9704079].\\
G.~P.~Korchemsky,
Nucl.\ Phys.\ B {\bf 498}, 68 (1997) [arXiv:hep-th/9609123];
arXiv:hep-ph/9801377.

\bibitem{gm}
A.~Gorsky and A.~Mironov,
arXiv:hep-th/0011197.
\bibitem{super}
P.~C.~Argyres, M.~R.~Plesser and A.~D.~Shapere,
Phys.\ Rev.\ Lett.\  {\bf 75} (1995) 1699 [arXiv:hep-th/9505100].\\
A.~Hanany and Y.~Oz,
Nucl.\ Phys.\ B {\bf 452}, 283 (1995) [arXiv:hep-th/9505075].\\
P.~C.~Argyres,
Adv.\ Theor.\ Math.\ Phys.\  {\bf 2}, 293 (1998)
[arXiv:hep-th/9706095]. \\
J.~A.~Minahan,
Nucl.\ Phys.\ B {\bf 537}, 243 (1999) [arXiv:hep-th/9806246].
\bibitem{kornew}
G.~P.~Korchemsky, J.~Kotanski and A.~N.~Manashov,
Phys. Rev. Lett. 88 (2002) 122002 [arXiv:hep-ph/0111185];
\\
  S.~E.~Derkachov, G.~P.~Korchemsky, J.~Kotanski and A.~N.~Manashov,
    hep-th/0204124.
\bibitem{khoze}
N.~Dorey, V.~V.~Khoze and M.~P.~Mattis,
Nucl.\ Phys.\ B {\bf 492}, 607 (1997) [arXiv:hep-th/9611016].
\bibitem{warner}
A.~Klemm, W.~Lerche, P.~Mayr, C.~Vafa and N.~P.~Warner,
Nucl.\ Phys.\ B {\bf 477} (1996) 746 [arXiv:hep-th/9604034].
\bibitem{witten}
E.~Witten,
Nucl.\ Phys.\ B {\bf 500}, 3 (1997) [arXiv:hep-th/9703166].
\bibitem{fgnr}
V.~Fock, A.~Gorsky, N.~Nekrasov and V.~Rubtsov,
JHEP {\bf 0007}, 028 (2000)
[arXiv:hep-th/9906235].

\bibitem{janik1}
R.~A.~Janik and J.~Wosiek,
Phys.\ Rev.\ Lett.\  {\bf 82}, 1092 (1999) [arXiv:hep-th/9802100].
\bibitem{KKM}
S.~E.~Derkachov, G.~P.~Korchemsky and A.~N.~Manashov,
Nucl.\ Phys.\ B {\bf 617} (2001) 375 [arXiv:hep-th/0107193].
\bibitem{devega}
H.~J.~De Vega and L.~N.~Lipatov,
Phys.\ Rev.\ D {\bf 64}, 114019 (2001) [arXiv:hep-ph/0107225].\\
H.~J.~de Vega and L.~N.~Lipatov,
arXiv:hep-ph/0204245.
\bi{sklyanin}
  E.K.~Sklyanin,
          Progr.\ Theor.\ Phys.\ Suppl.\ {\bf 118} (1995) 35 [solv-int/9504001].
\bibitem{gnr}
A.~Gorsky, N.~Nekrasov and V.~Rubtsov,
Commun.\ Math.\ Phys.\  {\bf 222}, 299 (2001)
[arXiv:hep-th/9901089].
\bibitem{witten2}
E. Witten, Nucl.\ Phys.\ B{\bf 323}, 113 (1989)

\end{thebibliography}
\end{document}